\documentclass{appolb}
\pdfoutput=1
\usepackage{blindtext}
\usepackage{typearea}
\usepackage{epsf}
\usepackage{graphics}
\usepackage{graphicx}
\usepackage{epstopdf}
\usepackage{caption}
\usepackage{subcaption}
\usepackage{amsmath}
\usepackage{slashed}
\usepackage{amssymb}
\usepackage{scalefnt}
\usepackage[normalem]{ulem}
\usepackage{cite}
\usepackage{etoolbox}
\usepackage{tikz}
\usepackage{color}
\usepackage{booktabs}
\usepackage[bottom]{footmisc}
\usepackage[ngerman,english]{babel}
\usepackage{microtype}
\usepackage [autostyle, english = american]{csquotes}
\MakeOuterQuote{"}
\usepackage{hyperref}
\usepackage{cleveref} 
\apptocmd{\thebibliography}{\setlength{\itemsep}{0.02cm}}{}{}
\Crefname{figure}{Fig.}{Figs.}

\newcommand{\citere}[1]{Ref.\,\cite{#1}}
\newcommand{\citeres}[1]{Refs.\,\cite{#1}}
\newcommand{\code}{\tt}

\newcommand{\sushi}[1]{{\code SusHi#1}}
\newcommand{\sushimi}[1]{{\code SusHiMi#1}}
\newcommand{\abbrev}{\scalefont{.9}}
\newcommand{\eqn}[1]{Eq.\,(\ref{#1})}
\newcommand{\fig}[1]{Fig.\,\ref{#1}}

\newcommand{\lhc}{{\abbrev LHC}}
\newcommand{\qcd}{{\abbrev QCD}}
\newcommand{\eft}{{\abbrev EFT}}
\newcommand{\sm}{{\abbrev SM}}

\newcommand{\mssm}{{\abbrev MSSM}}

\newcommand{\susy}{{\abbrev SUSY}}

\newcommand{\lo}{{\abbrev LO}}
\newcommand{\atlas}{{\abbrev ATLAS}}

\newcommand{\cp}{{\abbrev $\mathcal{CP}$}}
\newcommand{\nlo}{{\abbrev NLO}}
\newcommand{\nnlo}{{\abbrev NNLO}}
\newcommand{\nklo}[1]{{\abbrev N$^{#1}$LO}}

\newcommand{\mhmod}{m_h^{\mathrm{mod}+}}
\newcommand{\zhat}{\hat{\mathrm{\bf{Z}}}}

\newcommand{\td}[1]{\ensuremath{\tilde{#1}}}
\newcommand{\mc}[1]{\ensuremath{\mathcal{#1}}}

\newcommand{\tx}[1]{\ensuremath{\text{#1}}}
\newcommand{\h}[1]{\ensuremath{\hat{#1}}}
\newcommand{\nn}{\nonumber}
\newcommand{\al}{\alpha}
\newcommand{\bb}{\beta}
\newcommand{\D}{\Delta}

\pdfoutput=1
\usepackage{epsf,amsmath,amssymb,graphicx,dcolumn}
\usepackage{scalefnt,ulem,cite,hyperref}
\usepackage{cleveref,etoolbox,color}

\begin{document}
\title{\vspace*{-4em}
	\begin{flushright}
		{\sf\small
			KA-TP-02-2018\\
		}
	\end{flushright}
	\vspace*{2em}$\mathcal{CP}$-violating effects on MSSM Higgs searches}

\author{Shruti Patel$^{a,b}$, Elina Fuchs$^{c}$, Stefan Liebler$^{a}$, Georg Weiglein$^d$
\address{$^a$Institute for Theoretical Physics (ITP), Karlsruhe Institute of Technology,
	D-76131 Karlsruhe, Germany\\ 
	$^b$Institute  for  Nuclear  Physics  (IKP),  Karlsruhe  Institute  of  Technology, D-76344 Karlsruhe, Germany\\
	$^c$Department of Particle Physics and Astrophysics,
	Weizmann Institute of Science, Rehovot 76100, Israel\\
	$^d$DESY, Notkestra{\ss}e 85, D-22607 Hamburg, Germany\\[.3em]
	{\small\tt shruti.patel@kit.edu}\\[-.3em]
	{\small\tt elina.fuchs@weizmann.ac.il}\\[-.3em]
	{\small\tt stefan.liebler@kit.edu}\\[-.3em]
	{\small\tt georg.weiglein@desy.de}}\\
}
\maketitle
\begin{abstract}
We study the effects of \cp-violating phases on the phenomenology of the Higgs sector of the \mssm{}. Complex parameters in the \mssm{} lead to \cp-violating mixing between the tree-level \cp-even and \cp-odd neutral Higgs states, leading to three new loop-corrected mass eigenstates $h_a$, $a \in \lbrace 1,2,3\rbrace$. For scenarios where a light Higgs boson at about 125 GeV can be identified with the observed signal and where the other Higgs states are significantly heavier, a large admixture of the heavy neutral Higgs bosons occurs as a generic feature if \cp-violating effects are taken into account. Including interference contributions in the predictions for cross sections times branching ratios of the Higgs bosons is essential in this case.  As a first step, we present the gluon-fusion and bottom-quark annihilation cross sections for $h_a$ for the general case of arbitrary complex parameters, and we demonstrate that squark effects strongly depend on the phases of the complex parameters.  We then study the effects of interference between $h_2$ and $h_3$ for the example of the process $b\bar{b} \to \tau^+\tau^-$. We show that large destructive interference effects modify \lhc{} exclusion bounds such that parts of the parameter space that would be excluded by \mssm{} Higgs searches under the assumption of \cp-conservation open up when the possibility of \cp-violation in the Higgs sector is accounted for. 	

\end{abstract}
  
\section{Introduction}
Supersymmetric (\susy) models such as the Minimal Supersymmetric Standard Model (\mssm{}) or its next-to-minimal extension can not only alleviate many shortcomings of the Standard Model (\sm{}), but also accommodate the observed signal at 125 GeV~\cite{Aad:2012tfa,Chatrchyan:2012xdj} as one of several Higgs bosons predicted by their extended Higgs sectors.  
So far, the searches for additional Higgs bosons at the \lhc{} have been interpreted in various scenarios
beyond the \sm, including several supersymmetric ones.
However, the most general
case where \cp{} is violated and leads to mixing between \cp{}-even and -odd
eigenstates has not yet been covered by those analyses.
The reason for this has mainly been
the lack of appropriate theoretical predictions for the Higgs production
rates at the \lhc{} for the \cp-violating MSSM, and of a proper
prescription for taking into account relevant interference effects in Higgs
production and decay. In the following, we discuss state-of-the-art cross-section predictions in the \mssm{} for the two main Higgs production channels
at the \lhc{} which can be used as input for future experimental 
analyses in \cp{}-violating Higgs scenarios~\cite{Liebler:2016ceh,Patel:393346}. 
Additionally, an appropriate treatment of the interference effects arising in the calculation of cross section times branching ratio ($\sigma \times$BR) of a full process of production and decay of nearly mass-degenerate Higgs bosons is needed~\cite{Fowler:2010int,Fuchs:2015jwa,Fuchs:2017wkq,Fuchs:2016swt}. 
We review such a formalism, and subsequently study the implications of \cp{}-violating phases giving rise to Higgs mixing and interference in the process $b\bar{b} \to \tau^+\tau^-$. The resulting exclusion bounds are compared to existing experimental bounds from Run~II of the \lhc{}~\cite{Fuchs:2015jwa,Fuchs:2017wkq,Patel:393346}.

\section{The MSSM Higgs sector with complex parameters}\label{sec:higgs-mssm}
There are additional 105 free parameters in the \mssm{}, other than those from the \sm{}. These include $12$~physical, independent phases of the complex parameters of the \mssm{}. 
These phases are the ones of the soft-breaking gaugino masses $M_1$ and $M_3$, the Higgsino mass parameter $\mu$, and the trilinear soft-breaking
couplings~$A_f, f\in\lbrace e,\mu,\tau,u,d,c,s,t,b\rbrace$. The most restrictive constraints on the phases arise from bounds on
the electric dipole moments (EDMs) of the electron and the neutron
\cite{Demir:2003js,Chang:1998uc,Pilaftsis:1999td}. In the following discussion we focus on the phases $\phi_{A_t}$ and $\phi_{M_3}$, and their effects on the \mssm{} Higgs sector.

The \mssm{} contains two complex Higgs doublets with opposite hypercharges $Y_{\mathcal{H}_{1,2}} = \pm 1$
which induce masses for both the up- and down-type fermions.
The neutral fields of the two doublets
can be expressed in terms of \cp{}-even ($\phi_1^0, \phi_2^0$) and \cp{}--odd ($\chi_1^0, \chi_2^0$)
components as follows,
\begin{align}
\mathcal{H}_1 = \begin{pmatrix}
h_d^0 \\ h_d^-
\end{pmatrix} = 
\begin{pmatrix}
v_d + \frac{1}{\sqrt{2}}(\phi_1^0 + i \chi_1^0)\\
\phi^-_1
\end{pmatrix}\,, \quad
\mathcal{H}_2 = \begin{pmatrix}
h_u^+ \\ h_u^0
\end{pmatrix} = e^{i \xi}
\begin{pmatrix}
\phi^+_2\\
v_u + \frac{1}{\sqrt{2}}(\phi_2^0 + i \chi_2^0)	
\end{pmatrix} \label{eq:H_2}\,.
\end{align}
The two complex Higgs doublets possess eight degrees of freedom (dof). Three of these dof lend longitudinal components to the massive gauge bosons via the EWSB mechanism. The remaining physical dof manifest themselves as five Higgs bosons: \cp-even $h$ and $H$, \cp-odd $A$ and two charged Higgs states $H^{\pm}$.
The possible \cp-violating phase $\xi$ between the Higgs doublets vanishes at the minimum of the Higgs potential, and other possible phases in the tree-level Higgs potential can be rotated away. This makes the \mssm{} Higgs sector \cp-conserving at the lowest order. 
Besides the gauge couplings, it is fully determined by two parameters which are usually chosen as
$M_A$ or $M_{H^{\pm}}$  and $\tan \beta := \frac{v_u}{v_d}$.

\cp-violating effects enter the \mssm{} Higgs sector via radiative corrections. 
As a result of these \cp-violating loop effects, the tree-level mass eigenstates $\lbrace
h,H,A\rbrace$ mix into three \cp-admixed loop-corrected mass eigenstates $\lbrace h_1, h_2, h_3\rbrace$,
with $M_{h_1} \leq M_{h_2} \leq M_{h_3}$\footnote{The full mixing at higher orders takes place not just between $\lbrace h, H, A\rbrace$, but also with the Goldstone boson and the electroweak gauge bosons. Their impact is minimal for the processes considered here and therefore neglected in our treatment of the loop-corrected Higgs bosons, see \citeres{Liebler:2016ceh,Patel:393346} for a discussion.}.
In evaluating processes with external Higgs bosons beyond lowest order, an appropriate prescription to account for the mixings of the tree-level states into loop-corrected mass eigenstates is required so that the outgoing particle has the correct on-shell properties, and the S-matrix is properly normalised. This is established via the 
introduction of finite wave function normalisation factors, denoted as the so-called $\zhat$ factors~\cite{Chankowski:1992er,Dabelstein:1995js,Heinemeyer:2001iy}. The non-unitary $\zhat$ matrix contains the correction factors
for the external Higgs bosons $\lbrace h_1, h_2, h_3\rbrace$ relative to the lowest-order mass eigenstates $\lbrace h, H, A\rbrace$.
The matrix elements $\h{\mathbf{Z}}_{aj}$~\cite{Fuchs:2016swt,Williams:2007dc,Williams:2011bu}
are composed of the root of the external wave function
normalisation factor $\h{Z}_i^a$ and the on-shell transition ratio $\h{Z}^a_{ij}$,
\begin{align}
\h{Z}_i^a := \text{Res}_{\mc{M}_a^2} \lbrace \Delta_{ii} (p^2) \rbrace\,, \quad \h{Z}^a_{ij} = \frac{\Delta_{ij}(p^2)}{\Delta_{jj}(p^2)} \Bigg|_{p^2 = \mc{M}_a^2}\,,
\end{align}
which are evaluated at the complex pole $\mc{M}_a^2$.
Here $\Delta_{ij}$ are the propagators, $\lbrace a, b, c \rbrace$ denote the loop-corrected mass eigenstates, and
$\lbrace i, j, k \rbrace$ refer to the tree-level mass eigenstates.
With an appropriate assignment of the indices of the two types of states
(see \citere{Fuchs:2016swt}) the matrix elements can be written as
\begin{align}
\h{\mathbf{Z}}_{aj} = \sqrt{\h{Z}}_a \h{Z}_{aj}\,.
\end{align}
Using the $\zhat$ matrix elements, we obtain an expression for the amplitude of the loop-corrected mass eigenstates $h_a$ as a linear combination of the amplitudes of the tree-level states as follows\footnote{The ellipsis denote mixing contributions from the Goldstone bosons and electroweak gauge bosons which have been neglected.},
\begin{align}
\mathcal{A}_{h_a} = \zhat \begin{pmatrix}
\mc{A}_h \\ \mc{A}_H\\ \mc{A}_A
\end{pmatrix} = \zhat_{ah}\mathcal{A}_{h} + \zhat_{aH} \mathcal{A}_{H} + \zhat_{aA} \mathcal{A}_{A}+\ldots\,. \label{eq:eigenstate_mixing}
\end{align}

\section{Higgs production cross sections in the MSSM with complex parameters}\label{sec:higgsXS}

For low and medium values of $\tan \bb$ in the \mssm{}, Higgs bosons are predominantly produced through gluon fusion. At high $\tan \bb$, the production in association with a pair of bottom quarks is the dominant process, due to the enhanced bottom-Yukawa coupling to the Higgs bosons. In the following sections, we will present cross sections for production of Higgs bosons $h_1, h_2$ and $h_3$ in the \mssm{} via gluon fusion and bottom quark annihilation for a general case of arbitrary complex parameters. 

In the \mssm{}, the most significant contributions to the gluon-fusion cross section arise from top, bottom, stop and sbottom loops. Moreover, the weights of the top- and bottom-loop contributions have to be modified by the relative couplings to the \mssm{} Higgs bosons.  
For the case of the \mssm{} with complex parameters, \cp-violating phases can enter the cross section calculation via the $\zhat$ factors, Higgs--squark couplings, and through $\tan \beta$-resummed $\Delta_b$ corrections. These $\Delta_b$ corrections make the effective bottom Yukawa couplings explicitly complex, leading to $g_{b_L}^\phi \neq g_{b_R}^{\phi}$ (see~\citeres{Liebler:2016ceh,Patel:393346} for a discussion). Additionally, \cp-violating phases give rise to non-vanishing couplings of squarks to the \cp-odd state~$A$, $g^A_{\td{f}ii}$, which are zero when \cp{} is conserved.
\begin{figure}[t]
	\begin{center}
		\begin{tabular}{cc}
			\includegraphics[width=0.37\textwidth]{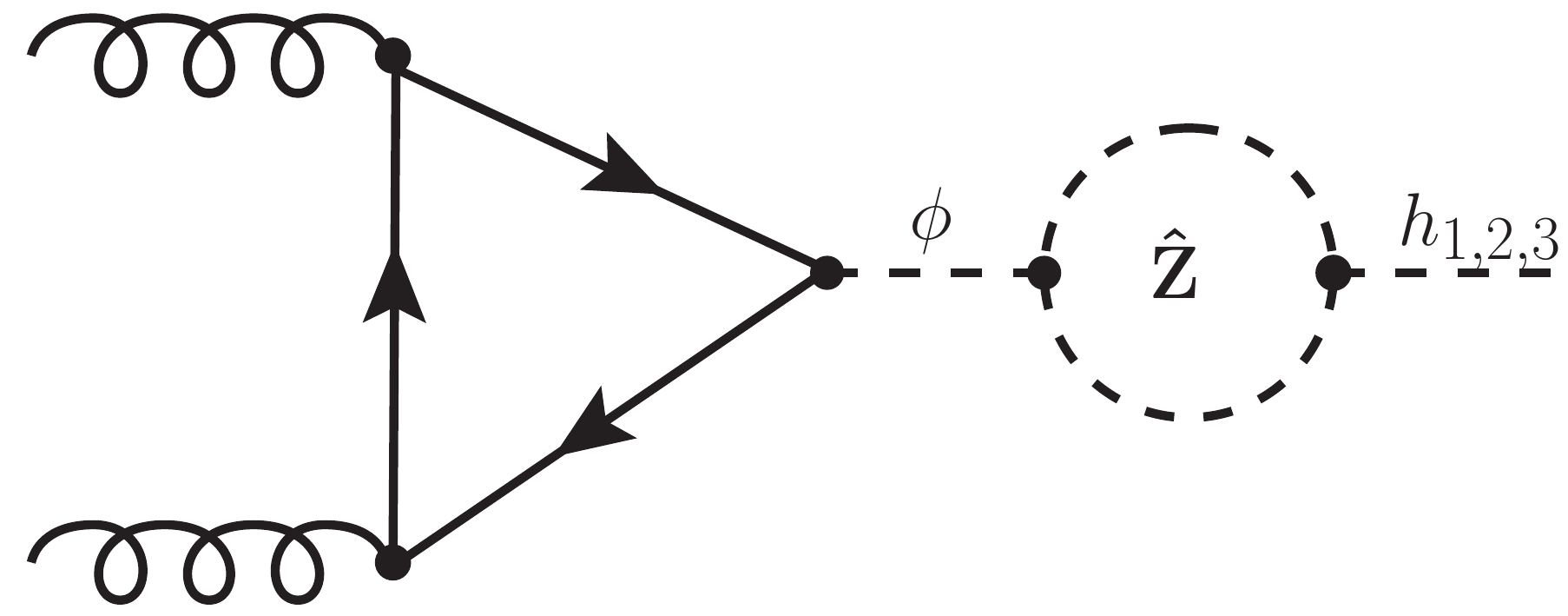}&
			\includegraphics[width=0.37\textwidth]{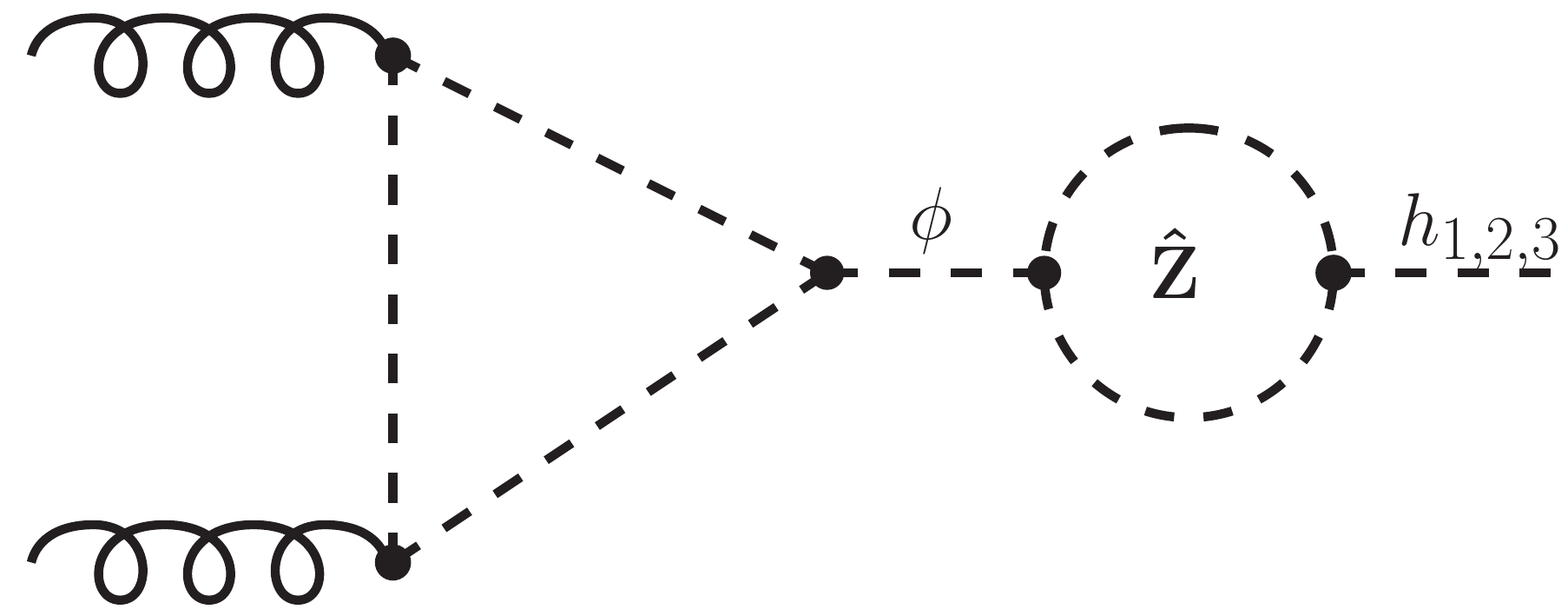}\\
			(a) & (b)
		\end{tabular}
	\end{center}
	\vspace{-0.3cm}
	\caption{Feynman diagrams for the
		\lo{} cross section with (a) quark and (b) squark contributions.}
	\label{fig: LOXS}
\end{figure}
The leading order (\lo{}) production cross section of the mass eigenstates~$h_a$ can be written as follows
\begin{align}
\label{eq:xs}
\sigma_{\text{\lo{}}}(pp\to h_a)=\sigma_0^{h_a}\tau_{h_a}\mathcal{L}^{gg}(\tau_{h_a})\quad \text{with}\quad
\mathcal{L}^{gg}(\tau)=\int_{\tau}^1\frac{dx}{x}g(x)g(\tau/x) \,,
\end{align}
where $\tau_{h_a}=m_{h_a}^2/s$. The hadronic squared 
centre-of-mass energy is denoted by $s$, and $\mathcal{L}^{gg}$ denotes the gluon--gluon luminosity. 
The partonic \lo{} cross section for $gg\to h_a$ is given by
\begin{align}
\label{eq:partonicxs}
\sigma_0^{h_a} = \frac{G_F \al_s^2}{288 \sqrt{\pi}}&\left[ \left|\mc{A}^{h_a,\mathrm{e}}  \right|^2 +
\left| \mc{A}^{h_a,\mathrm{o}}  \right|^2\right]\\\nonumber
\text{with} \quad &\mc{A}^{h_a,\mathrm{e}}=\zhat_{ah}\mc{A}_+^{h} + \zhat_{aH} \mc{A}_+^{H} + \zhat_{aA}  \mc{A}_-^{A}\\\nonumber
\text{and}\quad &\mc{A}^{h_a,\mathrm{o}}=\zhat_{ah}\mc{A}_-^{h} + \zhat_{aH} \mc{A}_-^{H} + \zhat_{aA}  \mc{A}_+^{A}\,,
\end{align}
where $G_F$ is Fermi's constant, and $\zhat_{a\phi}$ are the elements of the $\zhat$ matrix.
The "\lo{} cross section" is the cross section for the diagrams in Fig. \ref{fig: LOXS}  
despite the fact that it contains higher-order effects through the application of the $\zhat$~factors.
We see from \eqn{eq:partonicxs} that the final polarisation and colour averaged squared loop amplitude for a mass eigenstate $h_a$ consists of two non-interfering squared amplitudes. This is a because of the different tensor structures of various contributions. The amplitudes contributing to $\mc{A}^{h_a,e}$ have a symmetric tensor structure, while those contributing to $\mc{A}^{h_a,o}$ have an antisymmetric one. This results in the cross section being expressible as the sum of two non-interfering squared amplitudes.
This also explains the naming of the first and the second term with $\mc{A}^{h_a,\mathrm{e}}$, where $e$ denotes "even", and $\mc{A}^{h_a,\mathrm{o}}$, where $o$ denotes "odd", respectively.
Using this, we can split $\sigma_{\text{\lo{}}}$ into $\sigma_{\text{\lo{}}}^{\mathrm{e}}$ and $\sigma_{\text{\lo{}}}^{\mathrm{o}}$.

For the two \cp{}-even tree-level mass eigenstates $\phi^e\in \lbrace h,H\rbrace$ the amplitudes are
\begin{align}
\mc{A}_+^{\phi^e} = \sum_{q \in \lbrace t,b \rbrace} \left(a^{\phi^e}_{q,+} + \td{a}_{q}^{\phi^e}\right)\,,\quad  \mc{A}_-^{\phi^e} =  \sum_{q \in \lbrace t,b \rbrace} a^{\phi^e}_{q,-}\,.
\end{align}
Similarly, for the \cp{}-odd Higgs boson~$A$ the amplitudes are
\begin{align}
\mc{A}_-^{A} =  \sum_{q \in \lbrace t,b \rbrace} \left(a^{A}_{q,-} + \td{a}_{q}^{A}\right)\,,\quad \mc{A}_+^{A} =  \sum_{q \in \lbrace t,b \rbrace} a^{A}_{q,+}\,.
\end{align}
In the above expressions, $a^\phi_{q,+}$ and $a^\phi_{q,-}$ ($\phi \in \lbrace h,H,A \rbrace$) are the loop amplitudes for quark contributions proportional to the sum and difference of the right- and left-handed Yukawa couplings, respectively. The terms $\tilde{a}_q^\phi$ denote the loop amplitudes of squark contributions. The full expressions for $a_{q,\pm}^\phi$ and $\tilde{a}_q^\phi$ can be found in~\citere{Liebler:2016ceh}.  

The leading order cross section can be supplemented by higher-order corrections. At next-to-leading order (\nlo{}) the hadronic  cross section is given by the expression
\begin{align}
\label{eq:xsnlo}
\sigma_{\text{\nlo{}}}^{\mathrm{e/o}}(pp\to h_a+X)=\sigma_0^{h_a,\mathrm{e/o}}\tau_{h_a}\mathcal{L}^{gg}(\tau_{h_a})\left[1+C^{\mathrm{e/o}}\frac{\alpha_s}{\pi}\right]
+\Delta \sigma_{gg}^{\mathrm{e/o}}+\Delta\sigma_{gq}^{\mathrm{e/o}}+\Delta\sigma_{q\bar q}^{\mathrm{e/o}}\,.
\end{align}
The $\Delta\sigma$ terms contain the real corrections from the production of a Higgs boson in association with a gluon or quark jet. Note that the different left- and right-handed Yukawa couplings arise only for the case of the bottom quark due to the incorporation of the full $\Delta_b$ resummation, which makes the couplings explicitly complex at leading order~\cite{Williams:2011bu,Baglio:2013iia}. Beyond the leading order, we use a simplified $\D_b$ resummation which makes the left- and right-handed bottom Yukawa couplings equivalent to each other. Due to this, in the amplitudes for real corrections in the case of the \mssm{} with complex parameters the only new
ingredients, aside from the $\zhat$ factors, are Higgs--squark couplings $g_{\td{q}ii}^A$, which are added
to the \cp{}-even components~$\Delta\sigma^{\mathrm{e}}$. The real corrections can be split in
$\Delta\sigma^{\mathrm{e}}$ and $\Delta\sigma^{\mathrm{o}}$ since no interference terms arise. 

In the \mssm{} with real parameters, analytical \nlo{} virtual contributions
involving squarks, quarks and gluinos are either known in the
limit of a vanishing Higgs mass~\cite{Harlander:2003bb,Harlander:2004tp,Harlander:2005if,Degrassi:2008zj} or in an expansion of heavy \susy{} masses~\cite{Degrassi:2010eu,Degrassi:2011vq,Degrassi:2012vt}. In the \mssm{} with complex parameters the virtual contributions containing quarks have a similar structure as for those in the \mssm{} with real parameters, owing to the simplified $\D_b$ approximation beyond \lo{}. However, contributions from virtual corrections involving squarks are more difficult to generalise to complex parameters. We therefore interpolate these \nlo{} virtual contributions between phases $0$ and $\pi$
of the various \mssm{} parameters
using a cosine interpolation\cite{Heinemeyer:2006px,Hahn:2007fq}. For a certain value of the phase $\phi_z$ of a complex parameter $z$,
the virtual \nlo{} amplitude $\mc{A}_{\mathrm\nlo{}}^{\phi} (\phi_z)$
can be approximated using
\begin{align}\label{eq:cos_interpol}
\mc{A}_{\mathrm\nlo{}}^{\phi} (\phi_z) = \frac{1 + \cos \phi_z}{2}\mc{A}_{\mathrm\nlo{}}^{\phi} (0)
+ \frac{1 - \cos \phi_z}{2}\mc{A}_{\mathrm\nlo{}}^{\phi} ( \pi)
\end{align}
for each of the lowest-order mass eigenstates $\phi\in\lbrace h,H,A\rbrace$.
Here $\mc{A}_{\mathrm\nlo{}}^{\phi}(0)$ is the
analytical result for the \mssm{} with real parameters, and
$\mc{A}_{\mathrm\nlo{}}^{\phi}(\pi)$
is the analytical result with $z \rightarrow -z$. Finally, we also account for two-loop electroweak (EW) corrections mediated by light quarks, which are re-weighted to the \mssm{} with complex
parameters. The total gluon-fusion cross section at the $k$th order is the sum of the two parts
\begin{align}
\label{eq:ggphimaster}
\sigma_{\text{\nklo{k}}}(pp\to
h_a+X)=\sigma_{\text{\nklo{k}}}^{\mathrm{e}}(pp\to
h_a+X)+\sigma_{\text{\nklo{k}}}^{\mathrm{o}}(pp\to h_a+X)\,,
\end{align}
and the result beyond \lo{} \qcd{} is obtained through
\begin{align}
\label{eq:ggphimasternlo}
&\sigma^{\mathrm{e}}_{\text{\nklo{k}}} =
\sigma_{\text{\nlo{}}}^{\mathrm{e}}(1+\delta_{\mathrm{EW}}^{\mathrm{lf}})
+\left(\sigma_{\text{\nklo{k}, \text{\eft{}}}}^{t,\mathrm{e}} - \sigma_{\text{\nlo{}, \text{\eft{}}}}^{t,\mathrm{e}}\right)\\
&\sigma^{\mathrm{o}}_{\text{\nklo{k}}} =
\sigma_{\text{\nlo{}}}^{\mathrm{o}}
+\left(\sigma_{\text{\nklo{k}, \text{\eft{}}}}^{t,\mathrm{o}} -
\sigma_{\text{\nlo{}, \text{\eft{}}}}^{t,\mathrm{o}}\right) \,,
\end{align}
with $\delta_{\tx{EW}}^{\tx{lf}}$ containing the EW corrections from light fermions. 
\nklo{3} \qcd{} corrections are only taken into account for the \cp{}-even component
of the light Higgs boson, allowing us to match the precision of the light Higgs
boson cross section in the \sm{} used in up-to-date predictions. This
is because the light Higgs boson that is identified with the observed signal at 
$125\,\text{GeV}$ is usually assumed to have a dominant \cp{}-even
component, which is also the case in the scenarios which we consider in our numerical discussion.
For the \cp{}-odd component of the light Higgs and the heavy Higgs bosons we employ \nnlo{} corrections for the top-quark induced contributions in the effective theory of a heavy top-quark. This means that we do not account for
top-quark mass effects beyond \nlo{}, but only factor out the \lo{} \qcd{} cross
sections~$\sigma_{\text{\lo}}^{t,\mathrm{e}}$ and $\sigma_{\text{\lo}}^{t,\mathrm{o}}$.
These results have been implemented in an extension of the FORTRAN code \sushi{}~\cite{Harlander:2012pb,Harlander:2016hcx}, called \sushimi{} (SUsymmetric HIggs MIxing) and are currently the state-of-the-art for neutral Higgs
production in the \mssm{} with complex parameters~\cite{Liebler:2016ceh}. \sushimi{} will be included in the next release of \sushi{}.

Finally, for the production of the Higgs boson~$h_a$ via bottom-quark annihilation in the \mssm{} with complex
parameters, as implemented in \sushimi{}, the results for the \sm{}
Higgs boson are re-weighted to the \mssm{} with $|\zhat_{ah}g_b^h+\zhat_{aH}g_b^H|^2+|\zhat_{aA}g_b^A|^2$,
which includes $\tan\beta$-enhanced squark
effects through $\Delta_b$ using the simplified $\Delta_b$ resummation. 

We carry out our numerical analysis in a slightly modified version of the classic \mssm{} scenario introduced in \citere{Carena:2013ytb}, named the $\mhmod$ scenario. For this $\mhmod$-inspired scenario we choose for vanishing phases of the complex parameters:
\begin{align}\nonumber
&M_1=250\,\text{GeV},\quad M_2=500\,\text{GeV},\quad M_3=1.5\,\text{TeV}\\
&X_t=X_b=X_{\tau}=1.5\,\text{TeV},\quad A_q=A_l=0\\\nonumber
&\mu=\tilde{m}_{Q}=\tilde{m}_L=1\,\text{TeV}\,.
\end{align}
We use for the \sm{} parameters the values $m_t^{\textrm{OS}} = 173.20$\,GeV,
$m_b^{\overline{\textrm{MS}}}(m_b)=4.16\,$GeV,
$m_b^{\textrm{OS}}=4.75\,$GeV
and $\alpha_s (M_Z) = 0.119$. Furthermore, we choose $\tan\beta=10$
and $M_{H^{\pm}}=900$\,GeV. The $\mhmod$-inspired scenario features a lightest Higgs $h_1$ which is mostly \cp-even and \sm{}-like with a mass close to 125 GeV, and two heavier Higgs bosons $h_2$ and $h_3$ which are nearly mass-degenerate and heavily admixed. In our numerical analyses we employ {\tt{FeynHiggs-2.11.2}}~\cite{Heinemeyer:1998yj,Heinemeyer:1998np,Degrassi:2002fi,Frank:2006yh,Hahn:2013ria} to calculate Higgs masses and $\zhat$ factors. In the following, we study the variation of Higgs masses and cross sections with the phase of $A_t$. The phase $\phi_{A_t}$ is varied from $0$ to $2\pi$ leaving the absolute value $|A_t|$ constant in order to address various aspects in the phenomenology of Higgs boson production\footnote{We display the full range of the phase without imposing EDM constraints, following the common approach in studies of Higgs phenomenology with \cp-violation. See~\citere{Carena:2015uoe} for a recent discussion.}.  \begin{figure}[t!]
	\begin{center}
		\begin{tabular}{ccc}
			\includegraphics[width=0.3\textwidth]{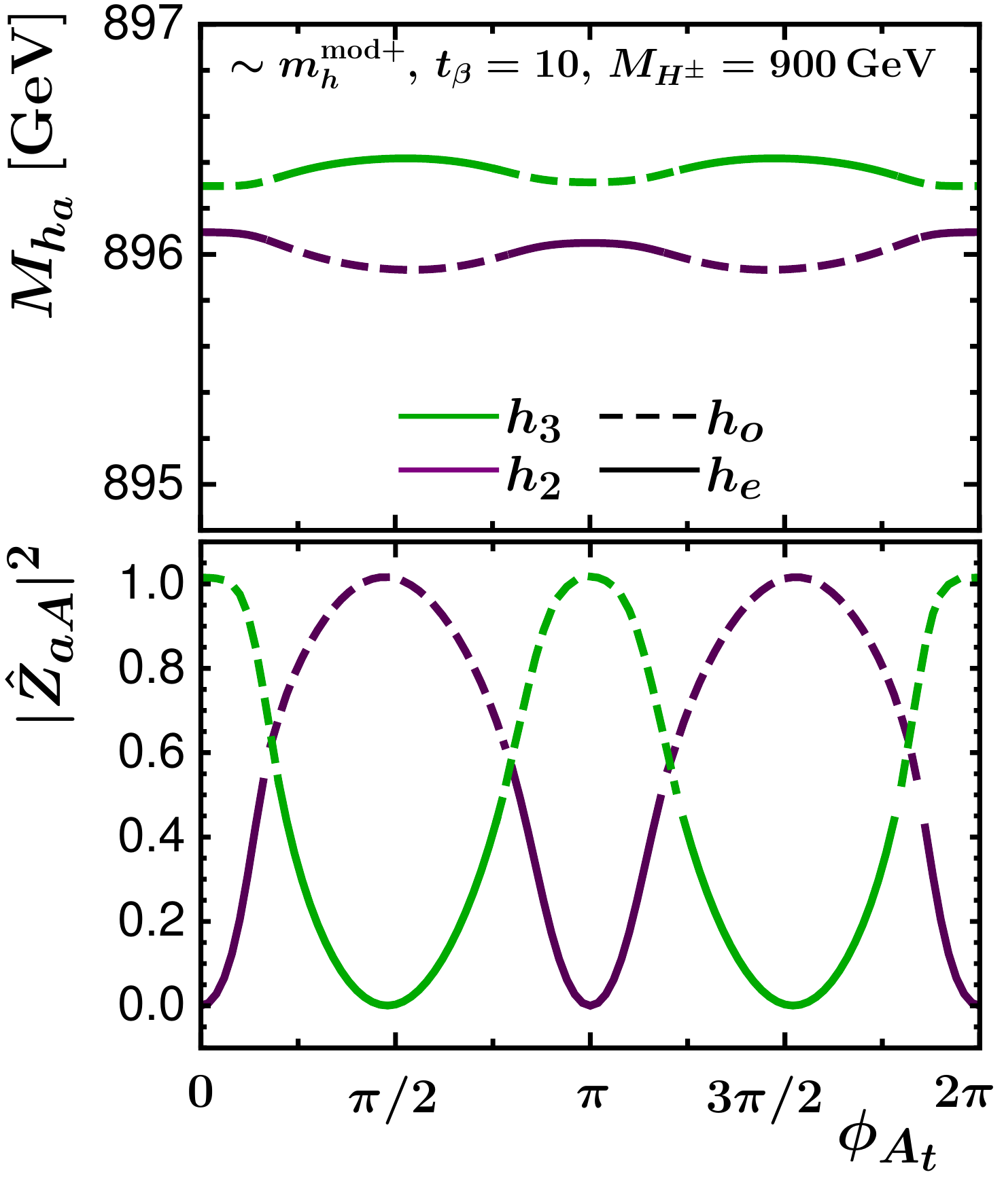} &
			\includegraphics[width=0.3\textwidth]{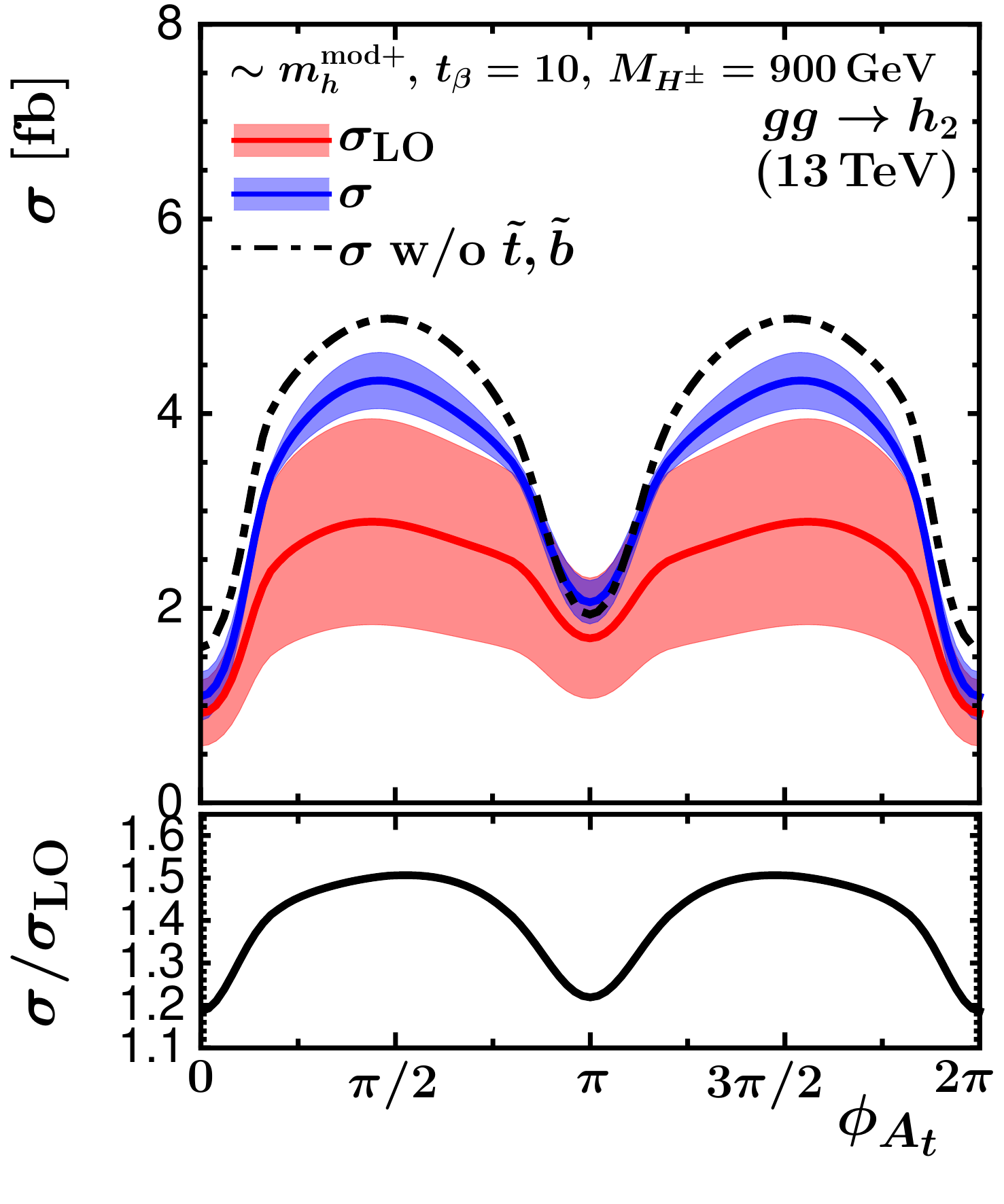} &
			\includegraphics[width=0.3\textwidth]{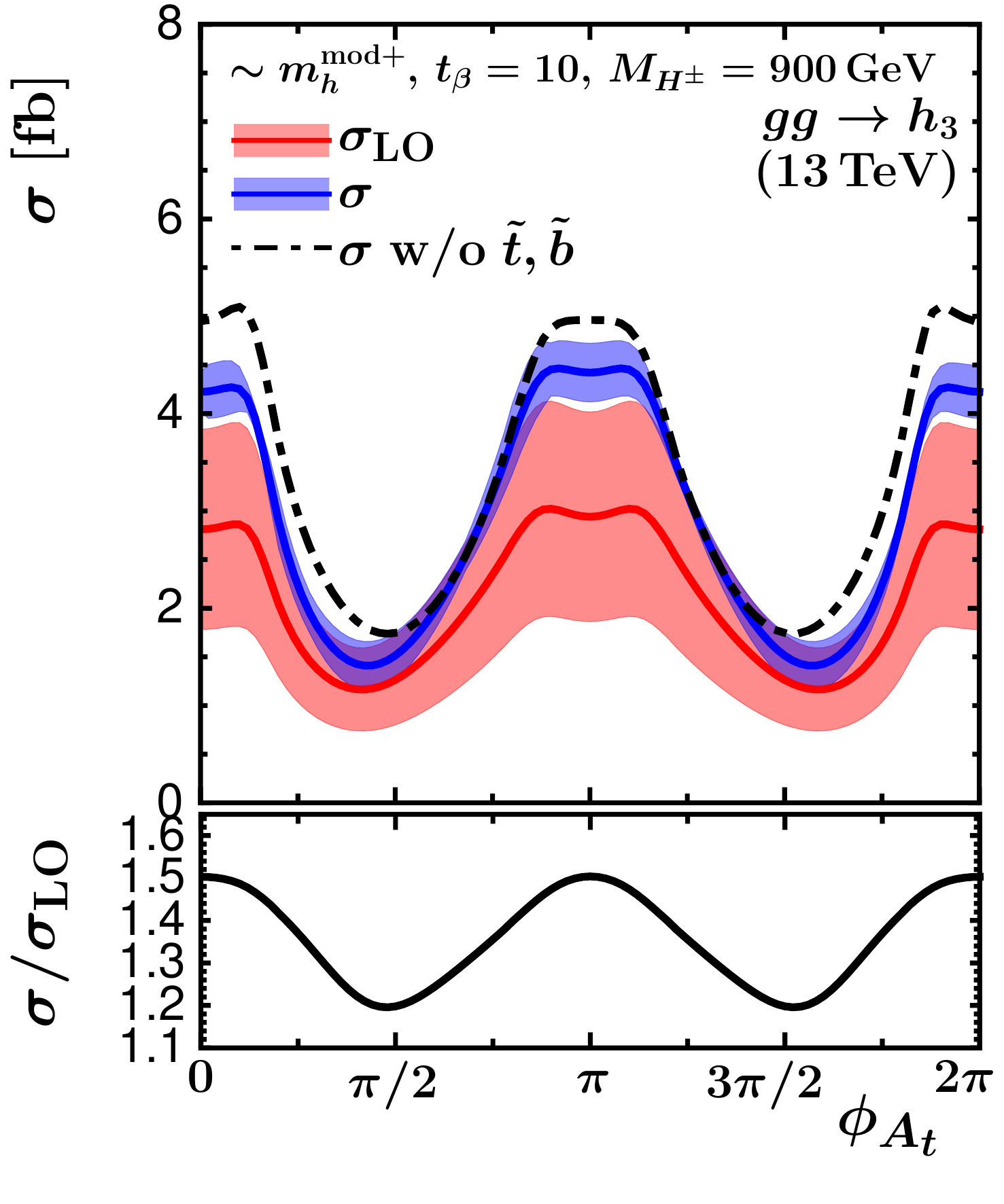} \\[-0.5cm]
			(a) & (b) & (c)\\
		\end{tabular}
	\end{center}
	\vspace{-0.2cm}
	\caption{Masses, mixing and gluon-fusion cross sections of $h_2$ and $h_3$ in the $\mhmod$-inspired scenario with $\tan \beta=10$. (a) Upper panel: Variation of $h_2$ (violet, lower curve) and $h_3$ (green, upper curve) masses in GeV with $\phi_{A_t}$.
		Lower panel: The \cp{}-odd character $|\zhat_{aA}|^2$ as a function of $\phi_{A_t}$. The solid and dashed curves represent
		regions in $\phi_{A_t}$ where $h_2$ and $h_3$ are predominantly
		\cp{}-even ($h_e$) or odd ($h_o$), respectively. At $\phi_{A_t}=0$ the state $h_3$ (green) is fully \cp-odd whereas $h_2$ (violet) is fully \cp-even. (b),(c) \lo{} (red, lowest curve) and best prediction (blue, middle curve) for the gluon-fusion cross section for (b) $h_2$ and (c) $h_3$ in fb
		as a function of $\phi_{A_t}$. The black, dot-dashed curve depicts the best prediction for the
		cross section without squark contributions (except through $\zhat$ factors). In the lower panel we show the $K$-factor $\sigma/\sigma_{\lo}$.}
	\label{fig:mhmod10}
\end{figure}

The variation of masses and \cp-character of the Higgs states $h_2$ and $h_3$ with $\phi_{A_t}$ is depicted in \fig{fig:mhmod10}~(a). We call the mass eigenstates
$h_2$ and $h_3$ either $h_e$ or $h_o$, depending on their mixing character:
if $|\zhat_{aA}|^2\gtrsim 1/2$ the mass eigenstate $h_a$ is denoted by $h_o$,
otherwise it is denoted by $h_e$. We see from the lower panel of \fig{fig:mhmod10}~(a) that while $h_3$ (in green) is fully \cp-odd
($|\zhat_{aA}|^2 \sim 1$)\footnote{Note that since the $\zhat$ matrix is non-unitary, its elements can have a value greater than 1.}  at $\phi_{A_t}=0$, and $h_2$ (in violet) is fully \cp-even ($|\zhat_{aA}|^2 \sim 0$) at $\phi_{A_t}=0$, their \cp-character varies widely as we scan through $\phi_{A_t}$, with both of them being substantially admixed for large parts of the $\phi_{A_t}$-space. \fig{fig:mhmod10}~(b) and (c) depict the cross sections for gluon fusion production of $h_2$ and $h_3$
in fb. The red curves with the larger  renormalisation and factorisation scale uncertainties associated with them show the variation of the \lo{} cross section with $\phi_{A_t}$. The blue curves with the reduced scale uncertainties show our best prediction cross section, whereas the black, dot-dashed curves depict the cross section with the squark contributions from loops turned off. Therefore, the only squark contributions to the black, dot-dashed curve come from the $\zhat$ factors. One notices that the blue (best prediction) curve and the dot-dashed curves follow each other closely and have a similar magnitude of the cross section. This implies that in this scenario, the phase dependence of the cross section comes mostly from the $\zhat$-factors, and not directly from squark loops. Moreover, the phase dependence of the cross sections closely follows the \cp-character of the Higgs states in \fig{fig:mhmod10}~(a). The lower panels of the cross section curves show the $K$-factors, which lie between about 1.2 and 1.5 with the phase 
$\phi_{A_t}$ and follow the variation the 
mixing character of $h_3$ and $h_3$.

Note that the curves for the cross sections of $h_2$ and $h_3$ have complementary shapes, and the two Higgs bosons are nearly mass degenerate. In such a case of nearly mass degenerate Higgs bosons, it may not be possible to experimentally resolve the two states as separate signals. The experimentally measured quantity would be the sum of the cross sections times their branching ratios along with the interference contributions in the full process of Higgs production and decay.  We will explore the effects of these \cp-violating interference contributions in the next section. 

\section{Impact of interference contributions}\label{sec:int}
At the \lhc{}, so far all searches for additional heavy Higgs bosons that have been interpreted in specific scenarios assume that the signal contributions from different Higgs bosons can be added incoherently, i.e.~without any interference effects, which is a valid assumption for the case of \cp{} conservation because the $h-H$ interference becomes large only in a small and already deeply excluded region of parameter space. However, if we allow for \cp{} violation, all three loop-corrected mass eigenstates $h_a, a \in \lbrace 1,2,3 \rbrace$ can interfere. Such interference effects are especially significant when the mass splitting between the Higgs bosons is smaller than the sum of their total widths, in which case the resonances can overlap. In order to accurately interpret the experimental limits on $\sigma \times$BR from Higgs searches at the \lhc{}, it is therefore crucial to also account for these interference contributions in their predictions, 
which could significantly enhance or diminish the value of $\sigma \times$BR in comparison to their values for the \cp{}-conserving case. 

We now consider a full process of Higgs production and decay and calculate the interference of amplitudes in an $s$-channel exchange of the Higgses $h_1, h_2$ and $h_3$ in a generic $2 \to 2$ parton level process
$I \to h_1, h_2, h_3 \to F$,
with the initial state $I$ denoting the production process and final state $F$ denoting the decay products. Later on, we will apply this to specific production and decay mechanisms. The calculation of the interference factors is carried out at leading order taking into account Higgs masses, total widths, and $\zhat$ factors from {\tt FeynHiggs-2.13.0} computed with full one-loop and leading two-loop contributions.
State-of-the-art higher-order contributions are taken into account in the computation for production cross sections for $I$ and branching ratios for $F$. For the \qcd{} corrections, a factorisation of higher-order corrections between initial and final states is often justified. This only misses corrections connecting initial and final state particles. Therefore it is well motivated to apply the interference factor calculated at \lo{} only, with the full process containing higher-order corrections. 

The interference term for a process $I \to F$ with $h_{1,2,3}$ Higgs exchange is obtained from the difference between the coherent and incoherent sum of the $2 \to 2$ squared amplitudes~\cite{Fuchs:2015jwa,Fuchs:2017wkq,Fuchs:2016swt}, 
\begin{align}
\label{eq:int-term}
|\mc{A}|^2_{\tx{int}} =&  \ |\mc{A}|^2_{\tx{coh}} - |\mc{A}|^2_{\tx{incoh}}\,,
\end{align}
where the coherent and incoherent sums are defined as
\begin{equation}
\label{eq:coh-incoh}
|\mc{A}|^2_{\tx{coh}}= \left|\sum\limits_{a=1}^3\mc{A}_{h_a}\right|^2 \,, \qquad |\mc{A}|^2_{\tx{incoh}}= \sum\limits_{a=1}^3\bigg|\mc{A}_{h_a}\bigg|^2\,.
\end{equation}
The squared amplitudes in \eqn{eq:int-term} and \eqn{eq:coh-incoh} can be used to define the cross sections $\sigma_{\text{int}}, \sigma_{\text{coh}}$ and $\sigma_{\text{incoh}}$. The relative interference term for the cross section of the full process is then defined as 
\begin{equation}
\label{eq:etaP}
\eta^{IF} = \frac{\sigma_{\textrm{int}}^{IF}}{\sigma_{\textrm{incoh}}^{IF}}\,.
\end{equation}
The total interference contribution to the process can be expressed as $\sigma_{\tx{int}} = \sigma_{\tx{int}_{12}} + \sigma_{\tx{int}_{23}} + \sigma_{\tx{int}_{13}}$, where $\sigma_{\tx{int}_{ab}}$ denotes the interference term between $h_a$ and $h_b$.
We then define the {relative} contribution for a \textit{single} Higgs $h_a$ from its interference with the Higgses $h_b$ and $h_c$ as
\begin{equation}\label{eq:intabac}
\eta_a^{IF} = \frac{\sigma^{IF}_{\textrm{int}_{ab}}}{\sigma^{IF}_{h_a} + \sigma^{IF}_{h_b}} + \frac{\sigma^{IF}_{\textrm{int}_{ac}}}{\sigma^{IF}_{h_a} + \sigma^{IF}_{h_c}}\,.
\end{equation}
Using $\eta_a^{IF}$ we can approximately factorise the experimentally measurable (coherent) cross section  as~~\cite{Fuchs:2015jwa,Fuchs:2017wkq,Fuchs:2016swt}
\begin{equation}\label{eq:xsbrint}
\sigma(pp \to I \to h_{1,2,3} \to F) \simeq \sum\limits_{a=1}^3 \sigma(pp \to I \to h_a) \cdot (1+\eta_a^{IF}) \cdot \mathrm{BR}(h_a \to F)\,.
\end{equation}
Currently \sushimi{} implements the relative interference factors for the heavy Higgs bosons, $\eta_2^{IF}$ and $\eta_3^{IF}$, for the case where only $h_2$ and $h_3$ interfere using \eqn{eq:intabac}~\cite{Patel:393346}.


In the following, we will study the effects of interference between $h_2$ and $h_3$ in the $b\bar{b} \to \tau^+ \tau^-$ process. For this purpose, we define a benchmark scenario, which we name \cp{}Int. 
Similar to the $\mhmod$-inspired scenario, the \cp{}Int scenario contains an \sm{}-like lightest Higgs, and two nearly mass degenerate and strongly admixed heavy Higgs bosons $h_2$ and $h_3$. Since $h_1$ is mostly \cp-even and has a large mass splitting from $h_{2,3}$, we only consider the interference between the two heavy Higgs bosons. The \cp{}Int scenario is defined with the following parameter values,
\begin{align}
&M_{\text{SUSY}}=1.5\,\text{TeV}\,,\quad \mu=1.5 \,\text{TeV}\,, \nn \\
&M_1=0.5\,\text{TeV},\quad M_2=1\,\text{TeV},\quad M_3=2.5\cdot e^{i\tfrac{\pi}{3}}\,\text{TeV}\,, \nn \\
&A_t=\left(\tfrac{\mu}{\tan \bb}+1.8\,M_{\text{SUSY}}\right)\cdot e^{i\tfrac{\pi}{4}},\quad A_b=A_t, \quad A_{\tau}=|A_t|\,, \nn \\ 
&M_{U_3}=M_{Q_3}=M_{D_3}=M_{\text{SUSY}}\,,\quad M_{L_{1,2}}=M_{E_{1,2}}=0.5\,\text{TeV}\,.
\end{align}
The \sm{} input parameters are $M_W=80.385$~GeV, $M_Z$ = $ 91.1876$~GeV,  $m_t^{\textrm{OS}} = 172.5$~GeV,
$m_b^{\overline{\textrm{MS}}}(m_b)$ = $4.18$~GeV, and $\alpha_s (m_Z)$ = 0.118, in accordance with the recommendations in~\citere{deFlorian:2016spz}. The 
phases of the parameters $A_t$ = $A_b$ and $M_3$ have been chosen to be non-maximal in view of the impact of bounds from EDMs. This benchmark illustrates the effects of mixing  and interference in $h_2, h_3$ production and decay. A more detailed study of the EDM constraints will follow in a forthcoming publication~\cite{MSSM2}.
\begin{figure}[t!]
	\begin{center}
		\begin{tabular}{cc}
			\includegraphics[width=0.43\textwidth]{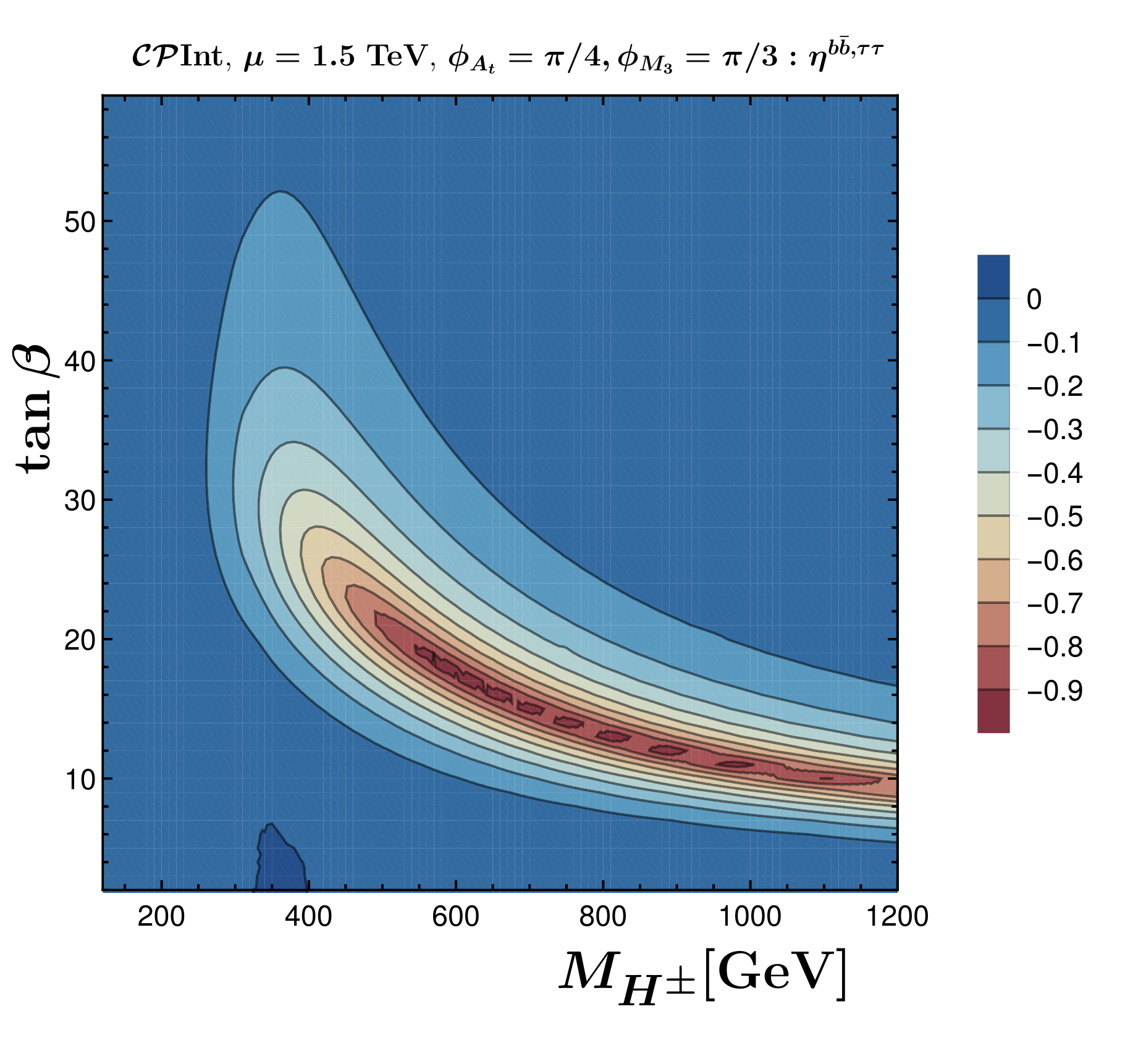} &
			\includegraphics[width=0.43\textwidth]{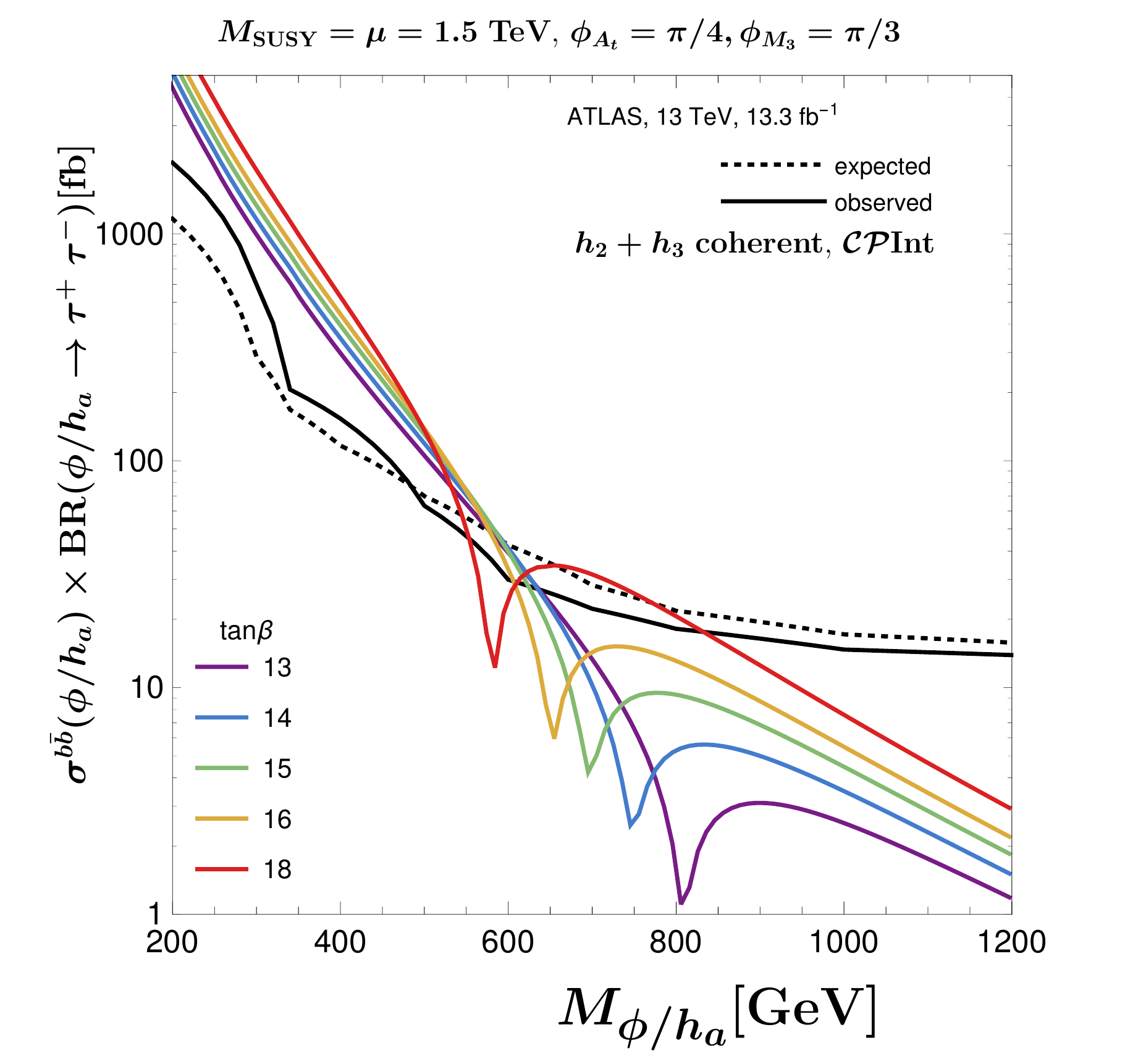} \\
			(a) & (b) \\
		\end{tabular}
	\end{center}
	\caption{Interference effects in the $b\bar{b} \to h_2, h_3 \to \tau^+\tau^-$ channel: (a) Contour plot for the interference factor $\eta^{b\bar{b},\tau\tau}$ in the $(M_{H^{\pm}}, \tan\beta)$ plane, and (b) comparison of  $\sigma \times$BR (in fb) for $h_2$ and $h_3$ including interference effects in the \cp{}Int scenario with the 95$~\%$ CL exclusion bounds obtained by \atlas{} at 13 TeV~\cite{ATLAS-CONF-2016-085}.}
	\label{fig:int}
\end{figure}

In \fig{fig:int}~(a) we show the relative interference factors $\eta^{b\bar{b},\tau\tau}$ for $b\bar{b}\to h_2,h_3\to \tau^+\tau^-$ in the $(M_{H^{\pm}}, \tan \beta)$ plane. 
As a result of the mass degeneracy between $h_2$ and $h_3$ and the fact that they are highly \cp-admixed, the interference contribution in their coherent $\sigma \times$BR is strongly destructive, with $\eta^{b\bar{b},\tau\tau}$ reaching a minimum of almost $-$98$\%$ in parts of the parameter space. This can be seen in \fig{fig:int}~(a), where we observe a valley of strong destructive interference of about $-$90$\%$ starting from around the points $(550\,\text{GeV}, 20)$ to $(1000\,\text{GeV}, 11)$ in the parameter plane. \fig{fig:int}~(b) depicts the theoretical predictions for $\sigma(pp \to b\bar{b} \to h_2, h_3 \to \tau^+\tau^-)$ as a function of the mass $M_{h_a}$ of a neutral scalar resonance $h_a$, along with the respective experimental limits for the production of a single resonance $\phi$ at mass $M_{\phi}$ obtained from \atlas{} searches for neutral Higgs bosons in Run~II at 13 TeV with $\int \mc{L}=13.3$ fb$^{-1}$ reported in~\citere{ATLAS-CONF-2016-085}. The black curves represent essentially model-independent upper limits on the $b\bar{b}$ production cross section times the  $\tau^+\tau^-$ branching ratio of a scalar boson versus its mass.  The solid black line represents the observed exclusion bound and the dotted black line depicts the expected bound. The theoretical predictions have been plotted for a sample of $\tan \bb$ values as a function of $M_{h_a}=M_{h_3}$, where in the relevant regions we also have $M_{h_3}\simeq M_{h_2}$. The top-most curve (in red) for the predicted $\sigma \times$BR corresponds to $\tan \beta=18$, while the bottom-most one (in violet) corresponds to $\tan \beta=13$. The comparison of the interference-corrected $\sigma \times$BR with the experimental limits can be understood as follows: the $M_{h_a}$ values corresponding to the parts of the predicted curves that lie \textit{above} the black experimentally measured curves are excluded at 95$\%$ CL, while those corresponding to the parts that lie below the experimental curves are still allowed. With this understanding, we can infer that for certain values of $\tan \beta$, the destructive interference suppresses the predicted $\sigma \times$BR below the experimental limits such that values of Higgs masses that would have been excluded if the interference contributions had not been taken into account are now allowed. 

\begin{figure}[t]
	\begin{center}
		\includegraphics[width=0.43\textwidth]{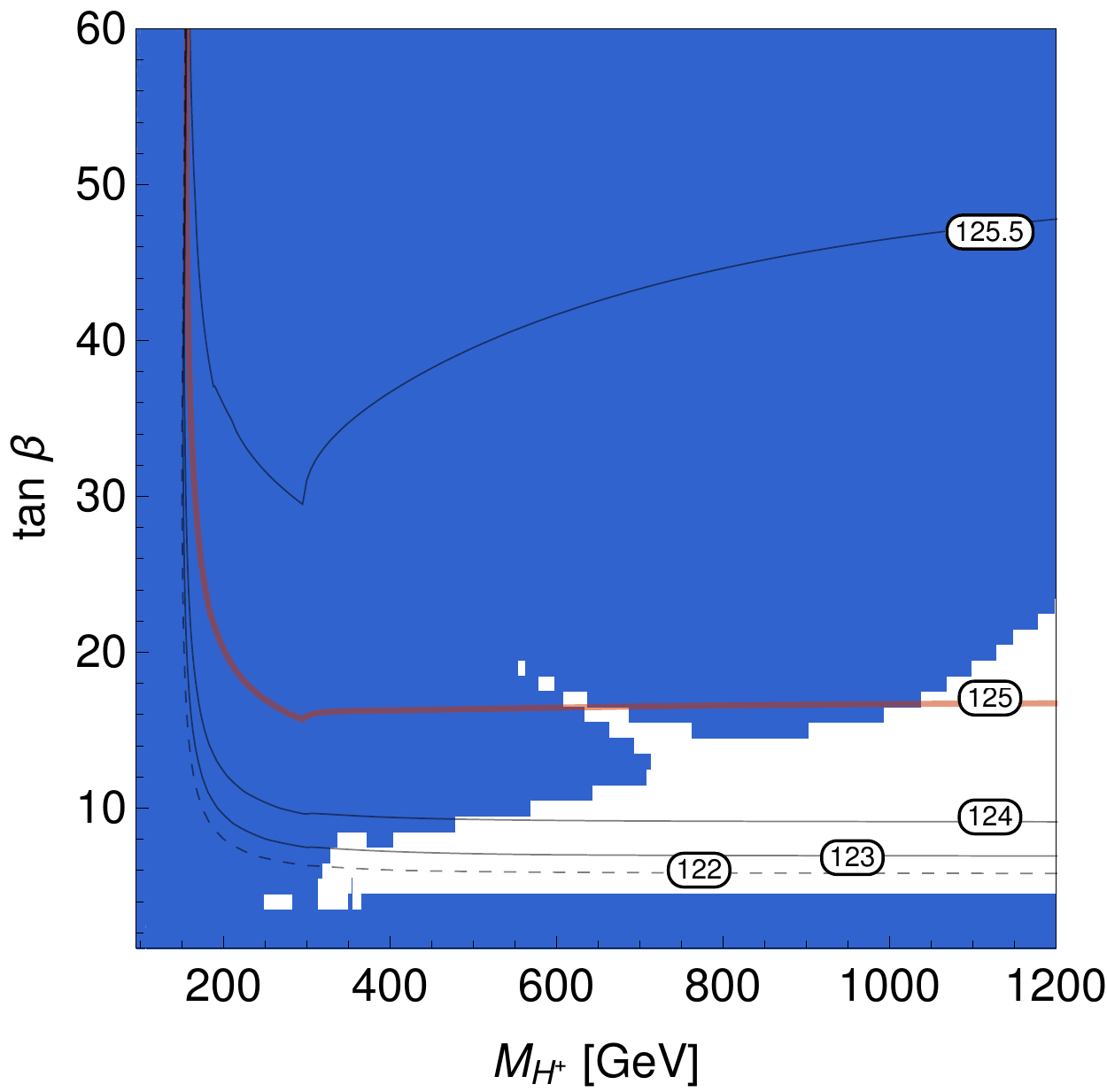} 
	\end{center}
	\caption{Exclusion bounds in the  $(M_{H^{\pm}}, \tan \bb)$ plane obtained with {\tt{HiggsBounds5.1.1beta}} for the \cp{}Int scenario. The blue region depicts the exclusion bounds when interference terms in the production and decay of $h_2$ and $h_3$ are taken into account for $b\bar{b} \to \tau^+\tau^-$. The contour lines depict the mass $m_{h_1}$ (in GeV) of the lightest Higgs boson.}
	\label{fig:HB}
\end{figure}

Finally, we analyse the exclusion limits in the $(M_{H^{\pm}}, \tan \bb)$ plane using the program {\tt{HiggsBounds-5.1.1beta}}\cite{Bechtle:2015pma, Bechtle:2013wla,Bechtle:2008jh,Bechtle:2013gu}. 
For any particular model, {\tt{HiggsBounds}} takes a selection of Higgs sector predictions as input and uses the experimental topological cross section limits from Higgs searches at LEP, the Tevatron and the \lhc{} to determine whether this parameter point has been excluded at 95$\%$ CL. In order to incorporate the interference effects into the prediction of $\sigma({b\bar{b} \to h_a})$ times the respective branching ratio, the ratio $\tfrac{\sigma^{\tx{model}}}{\sigma^{\tx{SM}}}$ of production cross sections which are used as input to {\tt{HiggsBounds}} are rescaled with the interference factor\footnote{In the new version of {\tt{HiggsBounds}} the interference factors can be directly given as an input.}. 
In \fig{fig:HB}, we show the modified exclusion bounds in the $(M_{H^{\pm}}, \tan \bb)$ plane, overlayed with $m_{h_1}$ contours. We see that accounting for the interference term and the complex parameters in the $\sigma \times$BR prediction leads to a "fjord" of destructive interference in the region between $M_{H^{\pm}} \sim 550$ GeV and 800 GeV for $\tan \bb \sim 13$ to 20 that remains unexcluded due to the suppression of the predicted $\sigma \times$BR. It is worthwhile to note that $m_{h_1} \sim 125$ GeV in this unexcluded space, making it a phenomenologically important region. 

\section{Conclusions}

Complex parameters in the \mssm{} give rise to rich and interesting phenomenology in the Higgs sector. Not only are such complex parameters needed for explaining the matter-antimatter asymmetry of the universe, they can also be extremely relevant for Higgs searches at the \lhc{}. We presented the full \lo{} cross section for gluon fusion supplemented with various higher-order contributions, and examined the three ways in which complex parameters affect the cross section, namely via $\zhat$ factors, complex Yukawa couplings due to $\Delta_b$ corrections, and Higgs--squark couplings.  
The bottom-quark annihilation cross section was treated with a simple re-weighting
procedure. 
Using the $\mhmod$-inspired scenario, we demonstrated the effects of \cp-violating Higgs mixing on the gluon-fusion Higgs production cross sections, and motivated the need to include interference contributions in the predictions for the $\sigma \times$BR of a full process of Higgs production and decay. Furthermore, we reviewed a formalism to consistently include such interference effects in our theoretical predictions and showed that taking into account \cp-violating mixing and interference contributions can significantly alter exclusion bounds from the \lhc{}. It is therefore essential to allow for the possibility that the \mssm{} Higgs sector may not be \cp-conserving when interpreting the latest data from \lhc{} Run II. 

\section*{Acknowledgements}
The authors acknowledge support by Deutsche Forschungsgemeinschaft through the SFB~676 ``Particles, Strings and the Early Universe''
and by the European Commission through the ``HiggsTools'' Initial Training Network PITN-GA-2012-316704.
\bibliography{literature}
\bibliographystyle{JHEP}
\end{document}